\documentclass[11pt,a4paper]{article}
\usepackage[english]{babel}
\usepackage{amsmath,amsthm,amssymb,epsfig,latexsym}
\usepackage{easybmat}
\usepackage{braket} 
\usepackage{color}
\usepackage{ulem}
\usepackage[numbers,square,sort&compress]{natbib}

\newcommand{\bu}{\bar u}
\newcommand{\bv}{\bar v}
\newcommand{\eps}{\epsilon}

\newcommand{\bw}{\bar w}

\newcommand{\cN}{\mathcal{N}_n^\nu(\bv)}

\newcommand{\be}[1]{\begin{equation}\label{#1}}
\newcommand{\ba}[1]{\begin{multline}\label{#1}}
\newcommand{\ee}{\end{equation}}
\newcommand{\ea}{\end{eqnarray}}

\newcommand{\tr}{\mathop{\rm tr}}

\newcommand{\Res}{\mathop{\rm Res}}

\newcommand{\dd}{\mathrm{d}}

\newcommand{\ZZf}{\mathbb{Z}_4}

\newtheorem{prop}{Proposition}[section]

\def\qed{\hfill\nobreak\hbox{$\square$}\par\medbreak}
 \makeatletter
 \@addtoreset{equation}{section}
 \makeatother

\newcommand{\bea}{\begin{eqnarray}}
\newcommand{\eea}{\end{eqnarray}}

\begin{document}

\vspace{12pt}

\begin{center}
\begin{LARGE}
{\bf   Form factor of local operators in the generalized algebraic Bethe ansatz}%
\end{LARGE}

\vspace{40pt}

{\large G.~Kulkarni\footnote{giridhar.kulkarni@ens-lyon.fr}} \\

Univ Lyon, ENS de Lyon, Univ Claude Bernard Lyon 1, CNRS, Laboratoire de Physique, F-69342 Lyon, France\\

\vspace{40pt}

{\large N.~A.~Slavnov\footnote{nslavnov@mi-ras.ru}}\\
Steklov Mathematical Institute of Russian Academy of Sciences, Moscow, Russia\\

 \vspace{12mm}

\end{center}

\vspace{1cm}


\begin{abstract}
We consider an $XYZ$ spin chain within the framework of the generalized algebraic Bethe ansatz. We study form factors of local operators corresponding to the
singlet states in the limit of free fermions. We obtain explicit representations for these form factors.
\end{abstract}

\vspace{4mm}


\vspace{1cm}

\section{Introduction}

In \cite{KulS23b}, we studied scalar products of Bethe vectors in the framework of the generalized algebraic Bethe ansatz \cite{FT79}. This work is a continuation of the paper \cite{KulS23b}.
Here we calculate form factors of local spin operators in an $XYZ$ Heisenberg chain.

A Hamiltonian of the periodic $XYZ$ chain \cite{Hei28} has the following form:
\be{HamXYZ}
H=\sum_{j=1}^{N} \Bigl (J_x \sigma^x_j\sigma^x_{j+1}+
J_y \sigma^y_j\sigma^y_{j+1}+J_z \sigma^z_j\sigma^z_{j+1}\Bigr ), \qquad  \sigma^{x,y,z}_{N+1}=\sigma^{x,y,z}_1.
\ee
It acts in a Hilbert space $\mathcal{H}=\mathcal{H}_1\otimes\mathcal{H}_2\otimes\cdots\otimes\mathcal{H}_N$, where each $\mathcal{H}_k\cong\mathbb{C}^2$.
The spin-$1/2$ operators $\sigma^{x,y,z}_k$ are Pauli matrices acting non-trivially in $\mathcal{H}_k$. Numerical coefficients  $J_{x,y,z}$
play the role of interaction strength along the axis $x$, $y$, and $z$. We assume that the number of sites $N$ is even.

To study the $XYZ$ chain, the generalized algebraic Bethe ansatz is used \cite{FT79}. This is because the $XYZ$ model is equivalent to an 8-vertex model \cite{Sut70,FanW70,Baxter71,Baxter-book} and has an 8-vertex $R$-matrix. However, when calculating the form factors of local operators, we use the same scheme as in the models solved by the standard algebraic Bethe ansatz \cite{FadST79,FadLH96,BogIK93L,Sla22L}. First, we express the local spin operators in terms of the monodromy matrix elements. This is achieved using the quantum inverse problem \cite{KitMT99,GohK00,MaiT00}. Second, we calculate the actions of the elements of the monodromy matrix on the Bethe vectors \cite{KulS23}. In the third stage, it is necessary to calculate the resulting scalar products
\cite{SlaZZ21,KulS23b}.  After performing the above steps, we obtain representations for the form factors of local operators in the form of some sums of scalar products. We investigate these representations in the present paper.

In the scheme described above, the last stage is the most difficult.
In models with a 6-vertex $R$-matrix, the scalar products of Bethe vectors  are well studied \cite{Sla89,KitMT99,BogIK93L,Sla22L}.  However, the scalar products  in models solvable by the generalized algebraic Bethe ansatz have been studied to a much lesser extent. To date, there is only one method that is based on reducing this problem to solving a system of linear equations \cite{BS19}. Using this method, determinant representations were obtained for scalar products, in which both vectors depend on the same number of parameters \cite{SlaZZ21}. Explicit representations for scalar products of a more general form have so far been obtained
in \cite{KulS23b} only for singlet states in the special case  when the $XYZ$ chain  is equivalent to free fermions ($XY$ model \cite{LieSM61}). Therefore, in this paper, we calculate the form factors of local operators only in this particular case.

The $XY$ model is well studied. Many different methods  allow one to calculate not only the form factors of local operators but also the correlation functions in this model \cite{LieSM61,McC68,Nie67,KatHS70,PerC77,VaiT78a,Ton81,DlorGS83,IzeKS98}. Therefore, we do not set ourselves the goal of obtaining new results. Our main goal is to show the fundamental applicability of the generalized algebraic Bethe ansatz to the calculation of the form factors. In the future, we plan to generalize this approach to the general case of the $XYZ$ chain.

The paper is organized as follows.  In section~\ref{S-N}, we give a brief description of the generalized algebraic Bethe ansatz.  In section~\ref{S-FF}, we define the form factors of local spin operators and describe their calculation scheme. In section~\ref{S-MR}, we present the main results of the paper. Finally, in section~\ref{S-ECFF}, we give examples of calculating the form factors. In particular, we show that singlet states have zero magnetization in section~\ref{SS-ZM}. In section~\ref{SS-TFF2}, we compute transversal form factors.

At the end of this paper we have collected basic information about Jacobi theta functions in  appendix~\ref{A-JTF}.
In appendix~\ref{A-CAF}, we describe numeric coefficients that arise when the monodromy matrix elements act on Bethe vectors. Finally, in appendix~\ref{A-SP}, we
give explicit representations for the scalar products of  Bethe vectors.

\section{Generalized algebraic Bethe ansatz for the XYZ model\label{S-N}}

In this section, we provide basic information about the description of the $XYZ$ model by the generalized algebraic Bethe ansatz.
The reader can get acquainted with this method in more detail in \cite{FT79,SlaZZ21}.

\subsection{$R$-matrix and monodromy matrix}

The  $R$-matrix  of the 8-vertex model has the following form:
\be{R-mat}
R(u)=\begin{pmatrix}
{\sf a}(u)&0&0&{\sf d}(u)\\
0&{\sf b}(u)&{\sf c}(u)&0\\
0&{\sf c}(u)&{\sf b}(u)&0\\
{\sf d}(u)&0&0&{\sf a}(u)\\
\end{pmatrix},
\ee
where
\be{abcd}
\begin{aligned}
&{\sf  a}(u)=\frac{2\theta_4(\eta |2\tau ) \, \theta_1(u+\eta |2\tau )\,
\theta_4(u|2\tau )}{\theta_2(0|\tau )\,\theta_4(0|2\tau )},
\\[8pt]
&{\sf  b}(u)=\frac{2\theta_4(\eta |2\tau ) \, \theta_4(u+\eta |2\tau )\,
\theta_1(u|2\tau )}{\theta_2(0|\tau )\,\theta_4(0|2\tau )},
\\[8pt]
&{\sf  c}(u)=\frac{2\theta_1(\eta |2\tau ) \, \theta_4(u+\eta |2\tau )\,
\theta_4(u|2\tau )}{\theta_2(0|\tau )\, \theta_4(0|2\tau )},
\\[8pt]
&{\sf  d}(u)=\frac{2\theta_1(\eta |2\tau ) \, \theta_1(u+\eta |2\tau )\,
\theta_1(u|2\tau )}{\theta_2(0|\tau )\, \theta_4(0|2\tau )}.
\end{aligned}
\ee
The definition of the Jacobi theta functions is given in appendix~\ref{A-JTF}.  The parameters $\eta$ and $\tau$ are related to the
coefficients $J_{x,y,z}$ (see \eqref{Jxyz}).

Within the framework of the generalized algebraic Bethe ansatz, Hamiltonian \eqref{HamXYZ} is constructed from the monodromy matrix $\mathcal{T}(u)$. This is a $2\times 2$ matrix, acting in an auxiliary space $\mathcal{H}_0\cong\mathbb{C}^2$
\be{Monod-def1}
\mathcal{T}(u)=\begin{pmatrix} A(u)& B(u)\\ C(u)& D(u)
\end{pmatrix}.
\ee
The matrix elements are operators depending on the complex $u$ and acting on the Hilbert space $\mathcal{H}$.
The monodromy matrix of the $XYZ$ chain of the length $N$ is equal to a product of the $R$-matrices acting in $\mathcal{H}_0\otimes\mathcal{H}_k$:
\be{Monod-def}
\mathcal{T}(u)=R_{01}(u-\xi_1)R_{02}(u-\xi_2)\cdots R_{0N}(u-\xi_N),
\ee
where complex parameters $\xi_k$ are called inhomogeneities. To construct the Hamiltonian \eqref{HamXYZ} we need only a homogeneous case  $\xi_k=0$,
$k=1,\dots,N$. Then
\be{Ham-TM}
H=    \frac{2\theta_1(\eta |\tau )}{\theta_1'(0|\tau )}  \frac{\dd}{\dd u}\log {\sf T}(u)\Bigr|_{u=0}
 -\frac{\theta_1'(\eta|\tau )}{\theta'_1(0 |\tau )} N \mathbf{1},
\ee
where $\mathbf{1}$ is the identity operator, and
\be{Transf-mat}
{\sf T}(u)={\tr} \mathcal{T}(u)=A(u)+D(u).
\ee
The operator ${\sf T}(u)$ is called a  transfer matrix. Eigenvectors of this operator coincide with the
ones of the Hamiltonian and other  integrals of motion.

The coupling constants of the Hamiltonian have the following representation:
\be{Jxyz}
J_x=\frac{\theta_4(\eta |\tau )}{\theta_4(0|\tau )}\,, \quad
J_y=\frac{\theta_3(\eta |\tau )}{\theta_3(0|\tau )}\,, \quad
J_z=\frac{\theta_2(\eta |\tau )}{\theta_2(0|\tau )}\,.
\ee
In the present paper, we focus on the case $\eta=1/2$, which corresponds to $J_z=0$. The corresponding model is called an $XY$ chain. It is equivalent to
free fermions.

Although only a homogeneous case is needed to construct the Hamiltonian of the $XYZ$ chain, in what follows we will consider a more general inhomogeneous model \eqref{Monod-def} with arbitrary complex inhomogeneities $\xi_k$. We emphasize, however, that we do this solely for reasons of generality. In all the formulas below, the homogeneous limit is trivial.

\subsection{Special notation}

Before moving on, we introduce some new notation. From now on, we  omit the modular parameter in the notation of theta functions
 whenever it is equal to $\tau$, namely, $\theta_a(\cdot)\equiv\theta_a(\cdot|\tau)$.

Let us also introduce two functions that will be often used below
\be{functions}
f(u,v)=\frac{\theta_1(u-v+\eta)}{\theta_1(u-v)},\qquad
h(u,v)=\frac{\theta_1(u-v+\eta)}{\theta_1(\eta)}. 
\ee

In what follows, we will constantly deal with sets of complex variables.
We  denote these sets by a bar: $\bu=\{u_1,\dots,u_m\}$, $\bv=\{v_1,\dots,v_n\}$, etc.  We also introduce special subsets $\bu_j=\bu\setminus\{u_j\}$, $\bu_{j,k}=\bu\setminus\{u_j,u_k\}$ and so on.

To make the formulas more compact we use a shorthand notation for products of  functions $f(u,v)$ and theta functions.
Namely, if the function  $f$ depends on a set (or two sets) of variables, this means that one should take the product over the corresponding set.
For example,
 \be{SH-prodllll}
f(u_j,\bu_j)=\prod_{\substack{u_l\in\bu\\ l\ne j}} f(u_j,u_l),  \quad f(\bv,\bu)=\prod_{\substack{u_l\in\bu\\ v_k\in\bv}} f(v_k,u_l)\qquad\text{etc.}
 \ee
Similarly,
 \be{SH-prodllllth}
\theta_2(u-\bv)=\prod_{v_k\in\bv} \theta_2(u-v_k),\qquad \theta_1(u_j-\bu_j)=\prod_{\substack{u_l\in\bu\\ l\ne j}} \theta_1(u_j-u_l),   \qquad\text{etc.}
 \ee
By definition, any product over the empty set is equal to $1$. A double product is equal to $1$ if at least one of the sets
is empty.

\subsubsection{Bethe vectors}

In this paper, the explicit form of  Bethe vectors is not essential. Therefore, we omit the details of their construction. The reader can find these details in
\cite{FT79,SlaZZ21}. We only note that the Bethe vectors are constructed using a special gauge transformation of the monodromy matrix.

We denote Bethe vectors by $|\hat\Psi^\nu_{n}(\bu)\rangle$. They belong to the Hilbert space $\mathcal{H}$: $|\hat\Psi^\nu_{n}(\bu)\rangle\in\mathcal{H}$. Bethe vectors are parameterized be a set of complex numbers $\bu=\{u_1,\dots,u_n\}$ and an integer $\nu$. In the case of free fermions, $\nu\in\ZZf$. If these parameters are related by a system of Bethe equations and a sum rule (see below), then this vector is an eigenvector of the transfer matrix. We call it an on-shell Bethe vector in this case. Otherwise, the Bethe vector is called off-shell.

Let us introduce
\be{chinu}
\chi_\nu(z)=(-1)^ne^{i\pi\nu/2}a(z)+e^{-i\pi\nu/2}d(z),
\ee
where
\be{ad-1}
a(z)=\theta_2 (z-\bar\xi),  \qquad d(z)= \theta_1 (z-\bar \xi).
\ee
Then Bethe equations have the following form:
\be{BE1}
\chi_\nu(u_j)=0,\qquad j=1,\dots,n.
\ee
Assume also that the parameters $\bu$ and $\nu$ satisfy the sum rule:
\be{sum-rule}
2\sum_{j=1}^nu_j=\sum_{k=1}^N\xi_k+\frac n2+\nu\tau+\nu_1,
\ee
where $\nu_1$ takes integer values.

If the conditions \eqref{BE1} and \eqref{sum-rule} are fulfilled, then
\be{actTvect}
{\sf T}(z)|\hat\Psi^\nu_n(\bu)\rangle=T_\nu(z|\bu)|\hat\Psi^\nu_n(\bu)\rangle,
\ee
where the transfer matrix eigenvalue $T_\nu(z|\bu)$ is
\be{Tnu1}
T_\nu(z|\bu)=\chi_\nu(z)f(z,\bu).
\ee

In the $XYZ$ model with a rational value of $\eta$, there is a degeneracy of the spectrum \cite{FMC03,FabM05,FabM04}. In particular, for $\eta=1/2$, the degeneracy is due to the presence of roots of Bethe equations differing from each other by $1/2$.
Let us define the following mapping over the fundamental domain $z\in\mathbb C/(\mathbb Z+\tau\mathbb Z)$
\begin{equation}
    z^\ast = z+\frac{(-1)^\eps}{2},
\end{equation}
where $\eps=0$ if $0\leq\Re z<\frac12$ and $\eps=1$ otherwise.
It is easy to check that $\chi_\nu(z^\ast)=(-1)^{\nu}\chi_\nu(z)$ (recall that we consider the chain of even length $N$). Therefore, if $u_j$ is a root $\chi_\nu(z)$, then $u_j^\ast$ is also a root $\chi_\nu(z)$. We will call $u_j$ and $u_j^\ast$ twins.

In what follows, we will work only with twin-free on-shell Bethe vectors that correspond to singlet eigenstates. Consideration of
vectors with twins requires a special study (see e.g. \cite{FabM06,Deg02,Deg02a,Fab07}).

For singlet states, $n=N/2$. In addition, we must require that there are no twins in the set $\bu$, that is, $u_j\ne u_k\pm1/2$ for any $u_j,u_k\in\bu$.

\begin{prop}
Let $\bu$ be a twin-free set of the roots of Bethe equations \eqref{BE1} satisfying the sum rule \eqref{sum-rule}. Then
\be{rep-chi}
\chi_\nu(z)=(-1)^ne^{-2\pi i\nu(z- \xi_p) +\pi i\nu/2}a(\xi_p)\frac{\theta_1(z-\bu)\theta_2(z-\bu)}{\theta_1(\xi_p-\bu)\theta_2(\xi_p-\bu)},
\ee
where $\xi_p$ is any of inhomogeneities.
\end{prop}

\textsl{Proof.}
Since equation \eqref{chinu} is an elliptic polynomial of degree $N$, it has $N$ roots in the fundamental domain. First, these are the roots
$\bu=\{u_1,\dots,u_n\}$. Second, these are their twins
$\bu^*=\{u^*_1,\dots,u^*_n\}$. 
Then
\be{prodth1}
\theta_1(z-\bu^*)=(-1)^{n'} \theta_2(z-\bu),
\ee
where $n'$ is the number of roots $u_j$ with $0\le \Re u_j<1/2$.

Consider a function
\be{fz}
{\sf f}(z)= \frac{\chi_\nu(z)}{\theta_1(z-\bu)\theta_1(z-\bu^*)}=\frac{(-1)^{n'}\chi_\nu(z)}{\theta_1(z-\bu)\theta_2(z-\bu)}.
\ee
Obviously, ${\sf f}(z)$ is an entire function of $z$.
Due to \eqref{A-shift} we have ${\sf f}(z+1)={\sf f}(z)$, and
\be{shift-tau1}
{\sf f}(z+\tau)=e^{-2\pi i\nu\tau}{\sf f}(z),
\ee
where we used  the sum rule.  Hence,
\be{fz2}
(-1)^{n'}\frac{e^{2\pi i\nu z} \chi_\nu(z)}{\theta_1(z-\bu)\theta_2(z-\bu)}=C,
\ee
where $C$ is a constant. To find this constant, we set $z=\xi_p$, where $\xi_p$ is an arbitrary inhomogeneity. Then
$d(\xi_p)=0$, and $\chi_\nu(\xi_p)=(-1)^n e^{i\pi\nu/2}a(\xi_p)$. Thus,
\be{const}
C=(-1)^{n'+n}\frac{e^{2\pi i\nu \xi_p+ i\pi \nu/2}a(\xi_p)}{\theta_1(\xi_p-\bu)\theta_2(\xi_p-\bu)},
\ee
and we arrive at \eqref{rep-chi}.\qed

The Bethe vectors also depend on two parameters of the gauge transformation $s$ and $t$, which are arbitrary complex numbers. We do not explicitly indicate this dependence in the notation.  In what follows, we will use the following combinations:
\be{xy}
x=\frac{s+t}2, \qquad y=\frac{s-t}2.
\ee
We also set
\be{sktk}
s_k=s+ \frac k2, \qquad t_k=t+ \frac k2, \qquad x_k=\frac{s_k+t_k}2.
\ee
For the singlet eigenstate, the dependence on $s$ and $t$ is present only in the scalar factor
\be{scal-fact}
|\hat\Psi^\nu_{n}(\bu)\rangle=\varphi(s,t)|\widetilde{\Psi}^\nu_{n}(\bu)\rangle,
\ee
where $|\widetilde{\Psi}^\nu_{n}(\bu)\rangle$ does not depend on the gauge parameters.
Therefore,  the dependence on $s$ and $t$ disappears in normalized expressions.

To calculate the form factors, we also need dual Bethe vectors $\langle\hat\Psi^\nu_{n}(\bu)|$. They belong to the dual space $\mathcal{H}^*$ and are arranged in a completely similar way to the Bethe vectors described above. In particular, they are left eigenvectors of the transfer matrix
\be{actTvectd}
\langle\hat\Psi^\nu_{n}(\bu)|{\sf T}(z)=T_\nu(z|\bu)\langle\hat\Psi^\nu_{n}(\bu)|,
\ee
if the parameters $\bu$ and $\nu$ satisfy conditions \eqref{BE1} and \eqref{sum-rule}. Then we call them dual on-shell Bethe vectors.

\section{Form factors\label{S-FF}}
In this section, we move on to the form factors of local operators, which is the main topic of this paper.
We call a form factor a matrix element of the form
\begin{equation}
    \mathcal F_{a,p}^{\nu,\lambda}(\bv|\bu)=
    \cN\braket{\hat\Psi^\nu_n(\bv)|\sigma^a_p|\hat\Psi^\lambda_n(\bu)},\qquad a\in\{x,y,z\}.
    \label{FFlocdef}
\end{equation}
Here $\langle\hat\Psi^\nu_n(\bv)|$ and $|\hat\Psi^\lambda_n(\bu)\rangle$ are, respectively, the left and right eigenstates of the transfer matrix. Recall that in this paper, we consider only the singlet part of the spectrum, that is, twin-free states. The normalization factor is chosen in a non-standard way
\be{Norm}
\cN=\left(\langle\hat\Psi^\nu_n(\bv)|\hat\Psi^\nu_n(\bv)\rangle\right)^{-1}.
\ee
This is because it was in this normalization that the scalar products were calculated in \cite{SlaZZ21,KulS23b}
\be{SPdef}
\mathbf S^{\nu,\mu}_{n,m}(\bv|\bw)=\cN\langle\hat\Psi^\nu_n(\bv)|\hat\Psi^\mu_m(\bw)\rangle.
\ee
Here $\langle\hat\Psi^\nu_n(\bv)|$ is an on-shell Bethe vector, while $|\hat\Psi^\mu_m(\bw)\rangle$ is an arbitrary off-shell vector.

Since the Bethe vectors in \eqref{FFlocdef} are not normalized, the form factor may depend on the parameters of the gauge transformation $s$ and $t$.
However, when calculating correlation functions, we usually deal with expressions that are quadratic in form factors, for instance,
\begin{equation}
    \mathcal F_{a,p}^{\nu,\lambda}(\bv|\bu)
    \mathcal F_{a,p'}^{\lambda,\nu}(\bu|\bv)=\cN\mathcal{N}_n^\lambda(\bu)\braket{\hat\Psi^\nu_n(\bv)|\sigma^a_p|\hat\Psi^\lambda_n(\bu)}
    \braket{\hat\Psi^\lambda_n(\bu)|\sigma^a_{p'}|\hat\Psi^\nu_n(\bv)}.
    \label{FFsqdef}
\end{equation}
We see that such expressions turn out to be normalized in the standard way. Therefore, they should not depend on the parameters of the gauge transformation. Such a disappearance of the dependence on $s$ and $t$ is one of the criteria for the correctness of the results obtained.

Using the solution of the inverse scattering problem\footnote{%
In \cite{KitMT99,MaiT00,GohK00}, the monodromy matrix was defined using the opposite ordering of the $R$-matrices. Therefore, formula \eqref{inv-sc-pr} is somewhat different from the solution of the quantum inverse problem given in the above-mentioned papers. } \cite{KitMT99,MaiT00,GohK00}
\be{inv-sc-pr}
\sigma_p^a=\left(\prod_{i=1}^{p}\sf T^{-1}(\xi_i)\right) \tr\big(\sigma^a\mathcal T(\xi_p)\big)  \left(\prod_{i=1}^{p-1}\sf T(\xi_i)\right),
\ee
we can reduce the form factors of local operators to the form factors of the monodromy matrix entries
\begin{equation}
    \mathcal F_{a,p}^{\nu,\lambda}(\bv|\bu)
    =\frac{\prod_{i=1}^{p-1}T_{\lambda}(\xi_i|\bu)}{\prod_{i=1}^{p}T_\nu(\xi_i|\bv)}
    \mathbf F_a^{\nu,\lambda}(\bv|\bu).
    \label{FFqism}
\end{equation}
Here $\mathbf F_a^{\nu,\lambda}(\bv|\bu)$ is represented as
\begin{equation}
    \mathbf F_a^{\nu,\lambda}(\bv|\bu)=
   \cN \braket{\hat\Psi^\nu_n(\bv)|\tr\big(\sigma^a\mathcal T(\xi_p)\big)|\hat\Psi^\lambda_n(\bu)}.
    \label{FFdef}
\end{equation}
We do not indicate dependence on $\xi_p$ in the notation $\mathbf F_a^{\nu,\lambda}(\bv|\bu)$ for brevity.

In its turn, the action of the operators $\tr\big(\sigma^a\mathcal T(\xi_p)\big)$ on Bethe vectors
$|\hat\Psi^\lambda_n(\bu)\rangle$  was calculated in \cite{KulS23}. For free fermions, it
has the following form:
\begin{multline}\label{actABCD-BV}
\tr\big(\sigma^a\mathcal T(w_{n+1})\big)|\hat\Psi^\lambda_n(\bw_{n+1})\rangle
=\sum_{\mu=0}^{3}\Bigg\{\sum_{k=1}^{n+1}
\mathbf{W}^{(\lambda,\mu)}_{a;0}(w_{n+1},w_k)|\hat\Psi^\mu_n(\bw_k)\rangle \\
+\sum_{k>l}^{n+1}
	\mathbf{W}^{(\lambda-\mu)}_{a;-}(w_{n+1},w_k,w_l)|\hat\Psi^\mu_{n-1}(\bw_{k,l})\rangle
+\mathbf{W}^{(\lambda-\mu)}_{a;+}(w_{n+1})|\hat\Psi^\mu_{n+1}(\bw)\rangle\Bigg\}.
\end{multline}
Here $w_{n+1}=\xi_p$ and $\bw_{n+1}=\bu$. The numerical coefficients $\mathbf{W}^{(\lambda,\mu)}_{a;0}$ and $\mathbf{W}^{(\lambda-\mu)}_{a;\pm}$ were calculated
in \cite{KulS23}. We give their explicit form in appendix~\ref{A-CAF}.

Equation \eqref{actABCD-BV} immediately allows us to reduce the form factors $\mathbf F_a^{\nu,\lambda}(\bv|\bu)$ to linear combinations of the scalar products
\begin{multline}\label{FF-linSP}
\mathbf F_a^{\nu,\lambda}(\bv|\bu)
=\sum_{\mu=0}^{3}\Bigg\{\sum_{k=1}^{n+1}
\mathbf{W}^{(\lambda,\mu)}_{a;0}(w_{n+1},w_k)   \mathbf S^{\nu,\mu}_{n,n}(\bv|\bw_k)\\
+\sum_{k>l}^{n+1}
	\mathbf{W}^{(\lambda-\mu)}_{a;-}(w_{n+1},w_k,w_l)\mathbf S^{\nu,\mu}_{n,n-1}(\bv|\bw_{k,l})
+\mathbf{W}^{(\lambda-\mu)}_{a;+}(w_{n+1})\mathbf S^{\nu,\mu}_{n,n+1}(\bv|\bw)\Bigg\}.
\end{multline}
The scalar products were calculated in \cite{SlaZZ21,KulS23b}. Their explicit form is given in appendix~\ref{A-SP}. Thus, the calculation of form factors is reduced to the substitution of known expressions into formula \eqref{FF-linSP}.

\section{Main results\label{S-MR}}

In this section, we present the main results of the paper.
Since the $XY$ model is planar, the longitudinal and transverse form factors are fundamentally different quantities. Therefore, we are considering
these two types of form factors separately.

\subsection{Longitudinal form factors\label{SS-LFF}}

Non-zero form factors $\mathcal F_{z,p}^{\nu,\lambda}(\bv|\bu)$ are those for which $\lambda=\nu$, $\nu=0,1$, and $\bu=\{\bv_k,v^*_k\}$, $k=1,\dots,n$. In other words, the set $\bu$ contains the twin of some element $v_k$, while all other elements are the same as elements of $\bv$. We can set $u_n=v^*_n$ and $\bu_n=\bv_n$
without loss of generality. Then
\begin{multline}\label{Obt3-res-ff}
\mathcal F_{z,p}^{\nu,\nu}(\bv|\{\bv_n,v^*_n\})=\left(\prod_{j=1}^{p-1}\frac1{f(\xi_j,v_n)f(v_n,\xi_j)}\right)\\
\times\frac{2(-1)^ne^{-i\pi\nu/2}\theta'_1(0)\theta_2(0)}{\mathcal{V}_n f(v_n,\bv_n)f(\bv_n,v_n)\theta_4(0|2\tau)}
\frac{\theta_2(v_n+y)}{\theta_1(v_n+y)}\frac{\theta_4(2v_n-2\xi_p|2\tau)}{\theta_2^2(v_n-\xi_p)},
\end{multline}
where
\be{logder}
\mathcal{V}_k=\frac{d}{dz}\log\frac{a(z)}{d(z)}\Big|_{z=v_k}.
\ee

We see that the individual form factor depends on the gauge transformation parameter $y$. However, this dependence disappears if we consider the quadratic
combination of form factors:
\begin{multline}\label{Obt3-res-ffff}
\mathcal F_{z,p}^{\nu,\nu}(\bv|\{\bv_n,v^*_n\})\mathcal F_{z,p'}^{\nu,\nu}(\{\bv_n,v^*_n\}|\bv)=\left(\prod_{j=p-1}^{p'-1}f(\xi_j,v_n)f(v_n,\xi_j)\right)\\
\times 4e^{-i\pi\nu}\left(\frac{\theta'_1(0)\theta_2(0)}{\mathcal{V}_n \theta_4(0|2\tau)}\right)^2
\frac{\theta_4(2v_n-2\xi_p|2\tau)\theta_4(2v_n-2\xi_{p'}|2\tau)}{\theta_2^2(v_n-\xi_p)\theta_1^2(v_n-\xi_{p'})},
\end{multline}
where we set $p'\ge p$ for definiteness.
\subsection{Transversal form factors\label{SS-TFF}}

Non-zero form factors  $\mathcal F_{x,p}^{\nu,\lambda}(\bv|\bu)$ and $\mathcal F_{y,p}^{\nu,\lambda}(\bv|\bu)$ occur for either $\nu=0$, $\lambda=1$
or $\nu=1$, $\lambda=0$. Then
\be{Two-FF1}
\begin{aligned}
&\mathcal F_{x,p}^{\nu,\lambda}(\bv|\bu)=-i\mu_0 \frac{i^{\mu_1}+(-i)^{\mu_1}}2\theta_4(0)\mathcal F_{p}^{\nu,\lambda}(\bv|\bu),\\
&\mathcal F_{y,p}^{\nu,\lambda}(\bv|\bu)=\frac{i^{\mu_1}-(-i)^{\mu_1}}2 \theta_3(0)\mathcal F_{p}^{\nu,\lambda}(\bv|\bu)
\end{aligned}
\ee
where
\be{Fcomm-res5x}
\mathcal F_{p}^{\nu,\lambda}(\bv|\bu)= \left(\prod_{j=1}^{p-1}\frac{f(\bu,\xi_j)}{f(\bv,\xi_j)}\right)
 e^{i\pi(\mu_0(s+\xi_p)-\tau/4+\nu/2)}S^\nu(\bv|\bu)\frac{\theta_2(0)}{\theta^2_4(0|2\tau)}
\frac{\theta_1(y+\bu)\theta_2(\xi_p-\bv)}{\theta_1(y+\bv)\theta_2(\xi_p-\bu)}.
\ee
In these formulas, $\mu_0=\lambda-\nu$, the function $S^\nu(\bv|\bu)$ is given by \eqref{S0}, and
\be{mu1}
\mu_1=2\sum_{j=1}^n(v_j-u_j)+\mu_0\tau.
\ee
Due to the sum rule, $\mu_1$ is an integer. Thus, the form factor $\mathcal F_{x,p}^{\nu,\lambda}$ vanishes for $\mu_1$ odd,
while $\mathcal F_{y,p}^{\nu,\lambda}$ vanishes for $\mu_1$ even.

We see that, as in the case above, we have a dependence on the gauge parameters. However, they are no longer included in the quadratic expressions. Indeed,
setting for definiteness $p'\ge p$ we obtain
\be{Two-FF2}
\begin{aligned}
&\mathcal F_{x,p}^{\nu,\lambda}(\bv|\bu)\mathcal F_{x,p'}^{\lambda,\nu}(\bu|\bv)= \frac{1+(-1)^{\mu_1}}2\theta_4^2(0)\mathfrak F_{p,p'}^{\nu,\lambda}(\bv|\bu),\\
&\mathcal F_{y,p}^{\nu,\lambda}(\bv|\bu)\mathcal F_{y,p'}^{\lambda,\nu}(\bu|\bv)=\frac{1-(-1)^{\mu_1}}2 \theta_3^2(0)\mathfrak F_{p,p'}^{\nu,\lambda}(\bv|\bu),
\end{aligned}
\ee
where
\begin{multline}\label{Fcomm-quadx}
\mathfrak F_{p,p'}^{\nu,\lambda}(\bv|\bu)
=-i\left(\prod_{j=p-1}^{p'-1}\frac{f(\bv,\xi_j)}{f(\bu,\xi_j)}\right) e^{i\pi\mu_0(\xi_p-\xi_{p'})-i\pi\tau/2}
\left(\frac{\theta_2(0)}{\theta^2_4(0|2\tau)}\right)^2 \\
\times S^\nu(\bv|\bu)S^\lambda(\bu|\bv)
\frac{\theta_2(\xi_p-\bv)\theta_2(\xi_{p'}-\bu)}{\theta_2(\xi_p-\bu)\theta_2(\xi_{p'}-\bv)},
\end{multline}
and we used $\nu+\lambda=1$.

\section{Examples of calculating form factors\label{S-ECFF}}

In this section, we give two examples of calculating form factors. In the first example, we prove that the magnetization of any singlet state is zero. In the second example, we consider the transversal form factor.

The calculation of form factors is straightforward but rather tedious. Therefore, we do not provide all the details. Instead, we focus only on the part of the computation that requires some non-trivial steps.

\subsection{Zero magnetization\label{SS-ZM}}

We begin our consideration with a magnetization
\begin{equation}
    \mathcal F_{z,p}^{\nu,\nu}(\bv|\bv)=
    \cN\braket{\hat\Psi^\nu_n(\bv)|\sigma^z_p|\hat\Psi^\nu_n(\bv)},\qquad p=1,\dots,N.
    \label{magnet}
\end{equation}
Due to \eqref{FFqism} we have
\begin{equation}
    \mathcal F_{z,p}^{\nu,\nu}(\bv|\bv)
    =\frac{1}{T_\nu(\xi_p|\bv)}
    \mathbf F_z^{\nu,\nu}(\bv|\bv),
    \label{magnetAD}
\end{equation}
where
\begin{equation}
    \mathbf F_z^{\nu,\nu}(\bv|\bv)=
   \cN \braket{\hat\Psi^\nu_n(\bv)|\big(A(\xi_p)-D(\xi_p)\big)|\hat\Psi^\nu_n(\bv)}.
    \label{magnetisp0}
\end{equation}
Recall that $\xi_p$ is  the inhomogeneity corresponding to the $p$th site.

In the process of calculations, the form factor $\mathbf F_z^{\nu,\nu+2}(\bv|\bv)$ also naturally arises. Therefore, from the very beginning, we will consider two form factors
\begin{equation}
    \mathbf F_z^{\nu,\lambda}(\bv|\bv)=
   \cN \braket{\hat\Psi^\nu_n(\bv)|\big(A(\xi_p)-D(\xi_p)\big)|\hat\Psi^\lambda_n(\bv)},\qquad \lambda=\nu,~\nu+2.
    \label{magnetisp}
\end{equation}
Note that, from a formal point of view, the Bethe equations \eqref{BE1} do not change when $\nu$ is replaced by $\nu+2$.

It is convenient to divide the form factor \eqref{magnetisp} into three parts, which correspond to the three terms on the rhs of \eqref{FF-linSP}:
\begin{equation}
   \mathbf F^{\nu,\lambda}_{z}(\bv|\bv)   =  \mathbf F^{\nu,\lambda}_{z,0}(\bv|\bv)
    +
    \mathbf F^{\nu,\lambda}_{z,+}(\bv|\bv)
    +
    \mathbf F^{\nu,\lambda}_{z,-}(\bv|\bv).
    \label{FFparts}
\end{equation}
We first consider $\mathbf F^{\nu,\lambda}_{z,0}(\bv|\bv)$. Using \eqref{Wa0} we obtain
\be{Fzz1}
\mathbf F^{\nu,\lambda}_{z,0}(\bv|\bv)
    =\sum_{\mu=0}^{3}\sum_{k=1}^{n+1}
\frac{\hat\alpha^z_{\lambda-\mu}(\xi_p-w_k)f(w_k,\bw_k)\Big( (-1)^ne^{i\pi\mu/2}a(w_k)-e^{-i\pi\mu/2}d(w_k)\Big)}
{4\theta_1(y+\xi_p)h(w_k,\xi_p)\theta_1(x)\theta_2(x)}
\mathbf S^{\nu,\mu}_{n,n}(\bv|\bw_k).
\ee
Here $\bw=\{\bv,\xi_p\}$ so that $w_{n+1}=\xi_p$ and $\bw_{n+1}=\bv$.

Due to selection rule \eqref{Ortho} $\mathbf S^{\nu,\nu+1}_{n,n}=\mathbf S^{\nu,\nu+3}_{n,n}=0$. Besides,
it is easy to show that $\hat\alpha_1^z(z)=\hat\alpha_3^z(z)=0$, because $\alpha^z_{l}(z)=\alpha^z_{l+2}(z)$. Thus, we conclude that
non-vanishing contributions to $\mathbf F^{\nu,\lambda}_{z,0}(\bv|\bv)$ occur only for $\lambda=\nu$ and $\lambda=\nu+2$. Therefore, we introduce the
following linear combinations:
\be{Feps}
\mathbf F^{\nu;\eps}_{z,0}(\bv|\bv)=\mathbf F^{\nu,\nu}_{z,0}(\bv|\bv)+(-1)^\eps\mathbf F^{\nu,\nu+2}_{z,0}(\bv|\bv), \qquad \eps=0,1.
\ee
Then it is easy to see that
\be{Fzz1-1}
\mathbf F^{\nu;\eps}_{z,0}(\bv|\bv)
    =\sum_{k=1}^{n+1}
\frac{\alpha^z_{\eps}(\xi_p-w_k)f(w_k,\bw_k)\Big( (-1)^ne^{i\pi\nu/2}a(w_k)-e^{-i\pi\nu/2}d(w_k)\Big)}
{\theta_1(y+\xi_p)h(w_k,\xi_p)\theta_1(x)\theta_2(x)}
\mathbf S^{\nu;1-\eps}_{n,n}(\bv|\bw_k),
\ee
where we used
\be{alplal}
\hat\alpha^z_{0}(z) + (-1)^\eps\hat\alpha^z_{2}(z) =4 \alpha^z_{\eps}(z).
\ee

One should distinguish two cases: $k=n+1$ and $k<n+1$. In the first case, $w_k=\xi_p$ and $\bw_k=\bv$. Then $d(w_k)=0$ and
$\mathbf S^{\nu;1-\eps}_{n,n}(\bv|\bv)=1$ (see \eqref{S0}--\eqref{Svu}). In the second case, $w_k=v_k$ and $\bw_k=\{\bv_k,\xi_p\}$.
Then
\be{adBE}
(-1)^ne^{i\pi\nu/2}a(v_k)-e^{-i\pi\nu/2}d(v_k)=2(-1)^ne^{i\pi\nu/2}a(v_k),
\ee
due to Bethe equations \eqref{BE1}, and
\be{SPP1}
\mathbf S^{\nu;1-\eps}_{n,n}(\bv|\{\bv_k,\xi_p\})=\frac{\theta'_1(0)a(\xi_p)f(\xi_p,\bv)\theta_2(v_k-\xi_p+x_\eps)}
{a(v_k)\mathcal{V}_k f(v_k,\bv_k)\theta_2(x_\eps)\theta_2(v_k-\xi_p)},
\ee
where $\mathcal{V}_k$ is given by \eqref{logder}.

Substituting these formulas into \eqref{Fzz1-1} we obtain
\be{Fzz2}
\mathbf F^{\nu;\eps}_{z,0}(\bv|\bv)
    =C_z\sum_{k=0}^{n}
    \mathcal{G}_{k;0}^z.
\ee
Here
\be{Fzcal}
C_z=\frac{e^{i\pi\nu/2}a(\xi_p)f(\bv,\xi_p)\theta'_1(0)\theta_2^2(0)}
{\theta_1(y+\xi_p)\theta_1(x_\eps)\theta_2^2(x_\eps)},
\ee
\be{Gk00z}
\mathcal{G}_{0;0}^z=\theta_2(y+\xi_p)\frac{\theta_1^2(x_\eps)\theta_2(x_\eps)}
{\theta'_1(0)\theta_2^2(0)},
\ee
and
\be{Gk0z}
\mathcal{G}_{k;0}^z=-2\frac{\theta_2(y+\xi_p)\theta_1(x_\eps)\theta_1(v_k-\xi_p-x_\eps)\theta_2(v_k-\xi_p+x_\eps)}
{\mathcal{V}_k\theta_1(v_k-\xi_p)\theta_2(v_k-\xi_p)\theta_2(0)}, \qquad k>0.
\ee

We do the same with the other two contributions to the form factor. Namely, we introduce
\be{Fepspm}
\mathbf F^{\nu;\eps}_{\pm,0}(\bv|\bv)=\mathbf F^{\nu,\nu}_{\pm,0}(\bv|\bv)+(-1)^\eps\mathbf F^{\nu,\nu+2}_{\pm,0}(\bv|\bv), \qquad \eps=0,1.
\ee
Then using \eqref{Wamin} and \eqref{Wapl} we obtain
\be{Fepsm1}
\mathbf F^{\nu;\eps}_{-,0}(\bv|\bv)=2i^\eps\sum_{k>l}^{n+1}
\frac{\beta^z_\eps(\xi_p,w_k,w_l)\omega_{kl}(\xi_p)}
{\theta_1(y+\xi_p)\theta_2(0)\theta^2_1(x)\theta^2_2(x)} \mathbf S^{\nu,\eps}_{n,n-1}(\bv|\bw_{k,l})
\ee
and
\be{Fepsp1}
\mathbf F^{\nu;\eps}_{+,0}(\bv|\bv)=-i^\eps\frac{\theta_2(0)\gamma^z_\eps(\xi_p)}{2\theta_1(y+\xi_p)}
\mathbf S^{\nu,\eps}_{n,n+1}(\bv|\bw),
\ee
where $\beta^z_\eps$,  $\gamma^z_\eps$, and $\omega_{kl}(z)$ respectively are given by \eqref{beta}, \eqref{gamma}, and \eqref{omg-dbl}. Recall also that $\bw=\{\bv,\xi_p\}$. To derive equations
\eqref{Fepsm1} and \eqref{Fepsp1}
we used $\beta_{l+2}=-\beta_l$ and $\gamma_{l+2}=-\gamma_l$, which results in
\be{betrgam}
\hat\beta_1+(-1)^\eps\hat\beta_3=4(-i)^\eps\beta_\eps,\qquad
\hat\gamma_1+(-1)^\eps\hat\gamma_3=4(-i)^\eps\gamma_\eps.
\ee
Using explicit representations for $\beta^z_\eps$ and the scalar product $\mathbf S^{\nu,\eps}_{n,n-1}$ we obtain
\be{Fzm2}
\mathbf F^{\nu;\eps}_{z,-}(\bv|\bv)
    =C_z\sum_{k=1}^{n}
    \mathcal{G}_{k;-}^z,
\ee
where
\be{Gkmz}
\mathcal{G}_{k;-}^z=\frac{\theta_1(t_\eps-\xi_p)\theta_1(v_k+s_\eps)\theta_1(v_k-\xi_p-x_\eps)}{\mathcal{V}_k \theta_2(v_k-\xi_p)\theta_1(v_k+y)}.
\ee

Similarly, using representations for $\gamma^z_\eps$ and the scalar product $\mathbf S^{\nu,\eps}_{n,n+1}$ we find
\be{Fzp2}
\mathbf F^{\nu;\eps}_{z,+}(\bv|\bv)
    =C_z\sum_{k=0}^{n}
    \mathcal{G}_{k;+}^z.
\ee
Here
\be{Gkpz0}
\mathcal{G}_{0;+}^z=e^{-i\pi\nu/2}
\frac{\theta_2(x_\eps) \theta_1(s_\eps+\xi_p)\theta_1(t_\eps-\xi_p)}{\theta'_1(0)\theta_2(y+\xi_p)}\frac{d(-y^*)}{\chi_\nu(-y^*)},
\ee
with $y^*=y-1/2$, and
\be{Gkpz}
\mathcal{G}_{k;+}^z=
\frac{\theta_1(s_\eps+\xi_p)\theta_1(v_k-t_\eps)\theta_2(v_k-\xi_p+x_\eps)}{\mathcal{V}_k \theta_1(v_k-\xi_p)\theta_2(v_k+y)},\qquad k>0.
\ee
Deriving contributions \eqref{Fzm2} and \eqref{Fzp2} we used the fact that $a(v_k)d(v_l)-a(v_l)d(v_k)=0$ due to the Bethe equations. Thanks to this property, all double sums that are initially present in the corresponding expressions turn into single ones.

Combining all the contributions above and using identities \eqref{Omm1}, \eqref{Omm2} we arrive at the following representation for the form factor:
\be{Fz3}
\mathbf F^{\nu;\eps}_{z}(\bv|\bv)
    =C_z\left\{\mathcal{G}_{0;0}^z+\mathcal{G}_{0;+}^z+\sum_{k=1}^{n}
    \mathcal{G}_{k}^z\right\},
\ee
where
\be{GCW-2}
\mathcal{G}_{k}^z= \frac{\theta_2(x_\eps)\theta_1(y+\xi_p)}{\theta_2(0)\mathcal{V}_k}\Big(\Omega(v_k-1/2)-\Omega(v_k)\Big),
\ee
and
\be{Ommmega}
\Omega(z)=\frac{\theta_2(z-\xi_p-x_\eps)\theta_2(z-\xi_p+x_\eps)\theta_1(z+y)}{\theta_1(z-\xi_p)\theta_2(z-\xi_p)\theta_2(z+y)}.
\ee
\textsl{Remark.} Obviously, $\Omega(z+1)=\Omega(z)$. Therefore, we can replace $\Omega(v_k-1/2)$ with $\Omega(v_k+1/2)$ in \eqref{GCW-2}. We have chosen the
minus sign for definiteness only.

Let us prove that the combination in braces in \eqref{Fz3}  vanishes. For this,
we consider a contour integral
\be{auxInt}
J=e^{-i\pi\nu/2}\frac{\theta_2(x_\eps)\theta_1(y+\xi_p)}{2\pi i\theta_2(0)}\oint\Omega(z)\frac{d(z)}{\chi_\nu(z)}\,\dd z,
\ee
where the integration is taken along the boundary of the fundamental domain. Due to the periodicity of the integrand, we conclude that $J=0$.
On the other hand, the integral is equal to the sum of the residues in the poles within the integration contour. First of all, we have poles in
the roots of Bethe equations $z=v_k$ and their twins $z=v_k^*$, $k=1,\dots,n$.  It is easy to see that these poles give us the sum of $\mathcal{G}_{k}^z$. There are two additional poles in
$z=-y+1/2=-y^*$ and $z=\xi_p+1/2$ (the pole at $z=\xi_p$ is compensated by the zero of the function $d(z)$, since $d(\xi_p)=0$). Thus, we obtain
\be{resInt}
\sum_{k=1}^n\mathcal{G}_{k}^z=-e^{-i\pi\nu/2}\frac{\theta_2(x_\eps)\theta_1(y+\xi_p)}{\theta_2(0)}
\Big(\Res\Omega(z)\frac{d(z)}{\chi_\nu(z)}\Bigr|_{z=-y^*}
+\Res\Omega(z)\frac{d(z)}{\chi_\nu(z)}\Bigr|_{z=\xi_p+1/2}\Big).
\ee
Evaluating the residues in the rhs of \eqref{resInt} we find
\be{resInt1}
\sum_{k=1}^n\mathcal{G}_{k}^z=-\mathcal{G}_{0;0}^z-\mathcal{G}_{0;+}^z.
\ee
Thus, we have proved that the form factor $\mathbf F^{\nu;\eps}_{z}(\bv|\bv)$ vanishes independently of the value of $\eps$. Hence,
$\mathbf F^{\nu,\nu}_{z}(\bv|\bv)=0$, which implies $\mathcal F_{z,p}^{\nu,\nu}(\bv|\bv)=0$.

\subsection{Transversal form factor\label{SS-TFF2}}

We consider the form factor $\mathbf{F}^{\nu,\lambda}_x(\bv|\bu)$ with $\nu=0$ and $\lambda=1$ for definiteness. As before, we introduce
\be{Fepspm-x}
\mathbf F^{0;\eps}_{x}(\bv|\bu)=\mathbf F^{0,1}_{x}(\bv|\bu)+(-1)^\eps\mathbf F^{0,3}_{x}(\bv|\bu), \qquad \eps=0,1.
\ee

Using the formulas of appendices~\ref{A-CAF} and~\ref{A-SP}, we obtain explicit expressions for all three contributions in formula \eqref{FF-linSP}. We
do not present the details of these calculations, since they are straightforward and completely analogous to the calculations of the previous section.
We only note that we use identities for theta functions of the form \eqref{con-thet1}, \eqref{GC}.

Ultimately, we present the form factor $\mathbf{F}^{\nu;\eps}(\bv|\bu)$ as follows:
\begin{equation}\label{Fcomm-res3}
\mathbf{F}^{\nu;\eps}(\bv|\bu)=H_\eps G_\eps.
\end{equation}
Here
\be{TXeps-1}
H_\eps=i^\epsilon a(\xi)f(\bu,\xi)S^0(\bv|\bu)
\frac{ \theta_4(0)\theta_2(0)}{2\theta_1(y+\xi)\theta_2(y+\xi)\theta_1(x)\theta_2(x)},
\ee
where $S^0(\bv|\bu)$ is given by  \eqref{S0}. The function $G_\eps$ has the following form:
\begin{multline}\label{Geps}
G_\eps=\theta_1(r+\xi_p+s_\eps)\theta_3(t_\eps-\xi_p)
\frac{\theta_2(y^*+\bu)\theta_2(\xi_p-\bv)}{\theta_2(y^*+\bv)\theta_2(\xi_p-\bu)}\\
+i\frac{\theta_1(\xi_p-\bu)\theta_1(y^*+\bv)}{\theta_1(\xi_p-\bv)\theta_1(y^*+\bu)} \theta_1(r-\xi_p+t_\eps)\theta_3(s_\eps+\xi_p)
\frac{\chi_1(-y^*)}{\chi_0(-y^*)},
\end{multline}
where $r=\sum_{j=1}^n(v_j-u_j)$.
Substituting here $\chi_1(-y^*)$ and $\chi_0(-y^*)$ in the form \eqref{rep-chi} we obtain
\be{Geps-tGeps}
G_\eps=\frac{\theta_2(y^*+\bu)\theta_2(\xi_p-\bv)}{\theta_2(y^*+\bv)\theta_2(\xi_p-\bu)}\tilde G_\eps,
\ee
where
\be{Geps1}
\tilde G_\eps=\theta_1(r+\xi_p+s_\eps)\theta_3(t_\eps-\xi_p)+
e^{2\pi i(y^*+ \xi_p)}
\theta_1(r-\xi_p+t_\eps)\theta_3(s_\eps+\xi_p).
\ee
Using
\be{thet-shift}
\theta_3(z)=\theta_2(z-\tfrac{\tau}2)e^{-i\pi(z-\tau/4)},
\ee
and $2y^*=s-t-1$ we transform \eqref{Geps1} as follows:
\be{Geps3}
\tilde G_\eps=e^{-i\pi(t_\eps-\xi_p-\tau/4)}\Big(\theta_1(r+\xi_p+s_\eps)\theta_2(t_\eps-\xi_p-\tfrac{\tau}2)-
\theta_1(r-\xi_p+t_\eps)\theta_2(s_\eps+\xi_p-\tfrac{\tau}2)\Big).
\ee
Observe that
\be{teps}
e^{-i\pi_0 t_\eps}=e^{-i\pi( t+\eps/2)}=(-i)^{\eps} e^{i\pi t}.
\ee
Thus, we obtain the following representation for the form factor \eqref{Fcomm-res3}:
\begin{equation}\label{Fcomm-res4}
\mathbf{F}^{0;\eps}(\bv|\bu)=H^{\text{mod}}_\eps G^{\text{mod}}_\eps.
\end{equation}
Here
\be{Xeps-mod}
H^{\text{mod}}_\eps= a(\xi_p)f(\bu,\xi_p)S^0(\bv|\bu)
\frac{e^{-i\pi(t-\xi_p-\tau/4)} \theta_4(0)\theta_2(0)}{2\theta_1(y+\xi_p)\theta_2(y+\xi_p)\theta_1(x)\theta_2(x)}
\frac{\theta_1(y+\bu)\theta_2(\xi_p-\bv)}{\theta_1(y+\bv)\theta_2(\xi_p-\bu)},
\ee
and
\be{Geps-mod}
G^{\text{mod}}_\eps=\theta_1(r+\xi_p+s_\eps)\theta_2(t_\eps-\xi_p-\tfrac{\tau}2)-
\theta_1(r-\xi_p+t_\eps)\theta_2(s_\eps+\xi_p-\tfrac{\tau}2).
\ee

Equation \eqref{Geps-mod} can be further simplified. Due to identity \eqref{the2-the1}, we obtain
\begin{multline}\label{Geps-mod2}
G^{\text{mod}}_\eps=(-1)^\eps\theta_1(r+2x-\tfrac{\tau}2|2\tau)\Big(\theta_4(2\xi_p+2y+r+\tfrac{\tau}2|2\tau)
-\theta_4(2\xi_p+2y-r-\tfrac{\tau}2|2\tau)\Big)\\
+\theta_4(r+2x-\tfrac{\tau}2|2\tau)\Big(\theta_1(2\xi_p+2y+r+\tfrac{\tau}2|2\tau)+\theta_1(2\xi_p+2y-r-\tfrac{\tau}2|2\tau)\Big),
\end{multline}
where we also used
\be{eps-theta}
\theta_1(2x_\eps+z|2\tau)=(-1)^\eps\theta_1(2x+z|2\tau),\qquad \theta_4(2x_\eps+z|2\tau)=\theta_4(2x+z|2\tau).
\ee

Due to the sum rule \eqref{sum-rule}, we have
\be{sum-rule999}
r=\tfrac{\mu_1}2-\tfrac{\tau}2,
\ee
where $\mu_1$ is an integer. Then
\begin{multline}\label{Geps-mod3}
G^{\text{mod}}_\eps=(-1)^\eps\theta_1(2x+\tfrac{\mu_1}2-\tau|2\tau)\Big(\theta_4(2\xi_p+2y+\tfrac{\mu_1}2|2\tau)
-\theta_4(2\xi_p+2y-\tfrac{\mu_1}2|2\tau)\Big)\\
+\theta_4(2x+\tfrac{\mu_1}2-\tau|2\tau)\Big(\theta_1(2\xi_p+2y+\tfrac{\mu_1}2|2\tau)
+\theta_1(2\xi_p+2y-\tfrac{\mu_1}2|2\tau)\Big).
\end{multline}
It remains to use
\be{diff-the}
\begin{aligned}
&\theta_4(z+\tfrac{\mu_1}2|2\tau)-\theta_4(z-\tfrac{\mu_1}2|2\tau)=0,\\
&\theta_1(z+\tfrac{\mu_1}2|2\tau)+\theta_1(z-\tfrac{\mu_1}2|2\tau)=\left(i^{\mu_1}+(-i)^{\mu_1}\right)\theta_1(z|2\tau).
\end{aligned}
\ee
This implies
\be{Geps-mod4}
G^{\text{mod}}_\eps=-i\left(i^{\mu_1}+(-i)^{\mu_1}\right)e^{i\pi(2x-\tau/2)}\theta_1(2x|2\tau)\theta_1(2\xi_p+2y|2\tau).
\ee
Substituting this into \eqref{Fcomm-res4} and using \eqref{the2-the1uu} we arrive at
\begin{multline}\label{Fcomm-res6}
\mathbf{F}^{0;\eps}(\bv|\bu)= -i\left(i^{\mu_1}+(-i)^{\mu_1}\right)a(\xi_p)f(\bu,\xi_p)S^0(\bv|\bu)\\
\times e^{i\pi(s+\xi_p-\tau/4)}\frac{ \theta_4(0)\theta_2(0)}{2\theta^2_4(0|2\tau)}
\frac{\theta_1(y+\bu)\theta_2(\xi_p-\bv)}{\theta_1(y+\bv)\theta_2(\xi_p-\bu)}.
\end{multline}
Since the result does not depend on $\eps$, we conclude that
\be{FF-fin-res}
\mathbf{F}^{0,1}(\bv|\bu)=\mathbf{F}^{0;\eps}(\bv|\bu), \qquad \mathbf{F}^{0,3}(\bv|\bu)=0.
\ee
Thus, we come to representation \eqref{Two-FF1}.

\section*{Conclusion}
In this paper, we have obtained explicit formulas for the form factors of local operators in the $XY$ model. We used for this the generalized algebraic Bethe ansatz
since the $XY$ model possesses the 8-vertex $R$-matrix. However, the general calculation scheme remains the same as when using the standard algebraic Bethe ansatz.
It includes the explicit solution of the quantum inverse problem, the calculation of the action of the monodromy matrix elements  on the Bethe vectors, and
the calculation of the resulting scalar products.

The last stage is the most technically difficult. That is why in this paper, we have limited ourselves to special cases of singlet states of the $XY$ chain.
In other cases, explicit representations are not yet known for all scalar products needed to compute the form factors. However, this obstacle is purely
technical. In \cite{KulS23b}, we described a method that allows one to obtain a system of linear equations for the scalar products of Bethe vectors in the case
of an arbitrary rational value of the parameter $\eta$. Having solved this system, we will be able to calculate the form factors of local operators in the
more general case of the $XYZ$ chain. We plan to address this issue in our future publications.

\section*{Acknowledgements}
We are grateful to A.~Zabrodin and A.~Zotov for numerous and fruitful discussions. The work of G.K. was supported by the SIMC postdoctoral grant of the Steklov
Mathematical Institute. Section~\ref{SS-TFF2} of the paper was performed by N.S.  The work of N.S. was supported by the Russian Science Foundation under grant
no.19-11-00062, https://rscf.ru/en/project/19-11-00062/ , and performed at Steklov Mathematical Institute of Russian Academy of Sciences.

\appendix

\section{Jacobi theta functions\label{A-JTF}}
Here we only give some basic properties of Jacobi theta functions used in the paper. See \cite{KZ15} for more details.

The Jacobi theta functions are defined as follows:
\be{JacTF}
\begin{aligned}
&\theta_1(u|\tau )=-i\sum_{k\in \mathbb{Z}}
(-1)^k q^{(k+\frac{1}{2})^2}e^{\pi i (2k+1)u},
\\[6pt]
&\theta_2(u|\tau )=\sum_{k\in \mathbb{Z}}
q^{(k+\frac{1}{2})^2}e^{\pi i (2k+1)u},
\\[6pt]
&\theta_3(u|\tau )=\sum_{k\in \mathbb{Z}}
q^{k^2}e^{2\pi i ku},
\\[6pt]
&\theta_4(u|\tau )=\sum_{k\in \mathbb{Z}}
(-1)^kq^{k^2}e^{2\pi i ku},
\end{aligned}
\ee
where $\tau \in \mathbb{C}$, $\Im \tau >0$, and
$q=e^{\pi i \tau}$.

Theta functions $\theta_a(u|\tau)$ with $a>1$ can be obtained from $\theta_1(u|\tau)$ by shifts of the argument
\be{A-shift0}
\begin{aligned}
&\theta_2(u|\tau)=\theta_1(u+\tfrac12|\tau),\\
&\theta_3(u|\tau)=e^{i\pi(u+\tau/4)}\theta_1(u+\tfrac12+\tfrac\tau2|\tau),\\
&\theta_4(u|\tau)=-ie^{i\pi(u+\tau/4)}\theta_1(u+\tfrac\tau2|\tau).
\end{aligned}
\ee
The following shift properties are important:
\be{A-shift}
\theta_1(u+1|\tau)=-\theta_1(u|\tau),\qquad \theta_1(u+\tau|\tau)=-e^{-\pi i(2u+\tau)}\theta_1(u|\tau).
\ee
Properties of $\theta_a(u|\tau)$ with $a>1$ with respect to the shifts by $1$ and $\tau$ follow from \eqref{A-shift0}.

Theta functions satisfy numerous identities based on periodicity \cite{KZ15}. We use some of them in this work. In particular, when
calculating the form factor of $\sigma_p^x$, we use
\be{con-thet1}
\frac{\theta_1(u+v)\theta_3(u-v)-\theta_1(u-v)\theta_3(u+v)}{\theta_2(u)\theta_4(u)}
=2\frac{\theta_1(v)\theta_3(v)}{\theta_2(0)\theta_4(0)}.
\ee
To prove \eqref{con-thet1}, it suffices to note that the lhs of this equation is a double periodic function of $u$ that has no poles in the fundamental domain. Therefore, this function is identically equal to a constant. Setting $u=0$, we arrive at \eqref{con-thet1}.

Similarly, one can prove an identity
\be{the4-the1}
2\theta_1(u+v|2\tau)\theta_4(u-v|2\tau)=\theta_1(u|\tau)\theta_2(v|\tau)+\theta_2(u|\tau)\theta_1(v|\tau).
\ee
It follows from this identity that
\be{the2-the1}
\theta_1(u|\tau)\theta_2(v|\tau)=\theta_1(u+v|2\tau)\theta_4(u-v|2\tau)+\theta_4(u+v|2\tau)\theta_1(u-v|2\tau).
\ee
In particular, setting $v=u$ in \eqref{the2-the1} we obtain
\be{the2-the1uu}
\theta_1(u|\tau)\theta_2(u|\tau)=\theta_1(2u|2\tau)\theta_4(0|2\tau).
\ee

It is often convenient to formulate the property of double periodicity as follows.
Let $\Phi(z)$ be a double periodic function with periods $1$ and $\tau$ and simple poles at $z=w_k$, $k=1,\dots,r$, in the fundamental domain. Then
\be{sumRes}
\sum_{k=1}^r\Res\Phi(z)\Big|_{z=w_k}=0.
\ee
This identity immediately follows from the fact that
\be{cint1}
\oint\Phi(z)\,\dd z=0,
\ee
where the integral is taken along the boundary of the fundamental domain.

In particular, when calculating the form factor of
$\sigma_p^z$, we use the following identities:
\begin{multline}\label{Omm1}
\theta_1(x_\eps)\theta_2(y+\xi_p)\theta_2(v_k+y)\theta_1(v_k-\xi_p-x_\eps)-
\theta_2(0)\theta_1(s_\eps+\xi_p)\theta_1(v_k-t_\eps)\theta_2(v_k-\xi_p)
\\
=-\theta_2(x_\eps)\theta_1(y+\xi_p)\theta_1(v_k+y)\theta_2(v_k-\xi_p-x_\eps),
\end{multline}
and
\begin{multline}\label{Omm2}
\theta_1(x_\eps)\theta_2(y+\xi_p)\theta_1(v_k+y)\theta_2(v_k-\xi_p+x_\eps)-
\theta_2(0)\theta_1(t_\eps-\xi_p)\theta_2(v_k+s_\eps)\theta_1(v_k-\xi_p)\\
=\theta_2(x_\eps)\theta_1(y+\xi_p)\theta_2(v_k+y)\theta_1(v_k-\xi_p+x_\eps).
\end{multline}
These identities respectively follow from
\be{cint-11}
\oint\frac{\theta_1(z+x_\eps)\theta_2(z+y+\xi_p)\theta_1(z+v_k-\xi_p-x_\eps)}{\theta_1(z)\theta_2(z)\theta_1(z+v_k+y)}\dd z=0,
\ee
and %
\be{cint-12}
\oint\frac{\theta_1(z-x_\eps)\theta_2(z+y+\xi_p)\theta_2(z+v_k-\xi_p+x_\eps)}{\theta_1(z)\theta_2(z)\theta_2(z+v_k+y)}\dd z=0.
\ee

Let us also give an example of a more sophisticated identity, which is used in the calculation of transversal form factors:
\begin{multline}\label{GC}
\sum_{k=1}^n
 \frac{\theta_1(u_k-\xi_p-x_\eps)\theta_2(r+u_k+s_\eps)}{\theta_2(u_k-\xi_p)\theta_1(u_k+y)}
\frac{\theta_1(u_k-\bv)}{\theta_1(u_k-\bu_k)}\\
=\frac{\theta_1(\xi_p+s_\eps)\theta_2(r+x_\eps)}{\theta_2(y+\xi_p)}
\frac{\theta_1(y+\bv)}{\theta_1(y+\bu)}
-\frac{\theta_2(x_\eps)\theta_1(r+\xi_p+s_\eps)}{\theta_2(\xi_p+y)}
\frac{\theta_2(\xi_p-\bv)}{\theta_2(\xi_p-\bu)}.
\end{multline}
Here $r=\sum_{j=1}^n(v_j-u_j)$. This identity follows from the analysis of a contour integral
\be{cint-3}
\oint \frac{\theta_1(z-\xi_p-x_\eps)\theta_2(r+z+s_\eps)}{\theta_2(z-\xi_p)\theta_1(z+y)}
\frac{\theta_1(z-\bv)}{\theta_1(z-\bu)}\,\dd z=0.
\ee
The sum of the residues in $z=u_k$, $k=1,\dots,n$, gives the lhs of \eqref{GC}. The residues in $z=\xi_p+1/2$ and $z=-y$ give
the rhs of \eqref{GC}.

.
\section{Coefficients of the action formula\label{A-CAF}}


To describe the coefficients $\mathbf{W}^{(\lambda,\mu)}_{a;0}(w_{n+1},w_k)$ we introduce three functions
\begin{equation}\label{alpha}
\begin{aligned}
    &\alpha^x_{l}(z)=(-1)^l\theta_4(y+w_{n+1})\theta_3(x_l)\theta_1(z+x_l),\\
    &\alpha^y_{l}(z)=i(-1)^l\theta_3(y+w_{n+1})\theta_4(x_l)\theta_1(z+x_l),\\
    &\alpha^z_{l}(z)=(-1)^l\theta_2(y+w_{n+1})\theta_1(x_l)\theta_1(z+x_l),
\end{aligned}
\end{equation}
and their Fourier transforms
\begin{equation} \label{four-alp}
\hat\alpha^a_\mu(z)=\sum_{l=0}^3e^{-i\pi\mu l/2}\alpha^a_l(z),\qquad a\in\{x,y,z\}.
\end{equation}
Then
\be{Wa0}
\mathbf{W}^{(\lambda,\mu)}_{a;0}(w_{n+1},w_k)=
\frac{\hat\alpha^a_{\lambda-\mu}(w_{n+1}-w_k)f(w_k,\bw_k)\Big((-1)^ne^{i\pi\mu/2}a(w_k)-e^{-i\pi\mu/2}d(w_k)\Big)}
{4\theta_1(y+w_{n+1})h(w_k,w_{n+1})\theta_1(x)\theta_2(x)}.
\ee

To describe the coefficients $\mathbf{W}^{(\lambda-\mu)}_{a;-}(w_{n+1},w_k,w_l)$ we introduce
three functions
\begin{equation}\label{beta}
\begin{aligned}
    &\beta^x_{l}(z,w_j,w_k)=\theta_4(0)\theta_3(t_l-z)\theta_1(z-w_j+x_l)\theta_1(z-w_k+x_l),\\
    &\beta^y_{l}(z,w_j,w_k)=i\theta_3(0)\theta_4(t_l-z)\theta_1(z-w_j+x_l)\theta_1(z-w_k+x_l),\\
    &\beta^z_{l}(z,w_j,w_k)=\theta_2(0)\theta_1(t_l-z)\theta_1(z-w_j+x_l)\theta_1(z-w_k+x_l),
\end{aligned}
\end{equation}
and their Fourier transforms
\begin{equation} \label{four-bet}
\hat\beta^a_\mu(z,w_j,w_k)=\sum_{l=0}^3e^{-i\pi\mu l/2}\beta^a_l(z,w_j,w_k),\qquad a\in\{x,y,z\}.
\end{equation}
Then
\be{Wamin}
\mathbf{W}^{(\lambda-\mu)}_{a;-}(w_{n+1},w_k,w_l)=\frac{\hat\beta^a_{\lambda-\mu}(w_{n+1},w_k,w_l)\omega_{kl}(w_{n+1})}
{2\theta_1(y+w_{n+1})\theta_2(0)\theta^2_1(x)\theta^2_2(x)},
\ee
where
\begin{equation}
    \omega_{kl}(z)=
    \frac{\left[d(w_k)a(w_l)-d(w_l)a(w_k)\right]f(w_k,\bw_k)f(\bw_l,w_l)}{f(w_k,w_l)h(w_k,z)h(z,w_l)}.
    \label{omg-dbl}
\end{equation}

Finally, to describe the coefficients $\mathbf{W}^{(\lambda-\mu)}_{a;-}(w_{n+1})$ we introduce three functions
\begin{equation}\label{gamma}
\begin{aligned}
    &\gamma^x_{l}(z)=\theta_4(0)\theta_3(s_l+z),\\
    &\gamma^y_{l}(z)=i\theta_3(0)\theta_4(s_l+z),\\
    &\gamma^z_{l}(z)=\theta_2(0)\theta_1(s_l+z),
\end{aligned}
\end{equation}
and their Fourier transforms
\begin{equation} \label{four-gam}
\hat\gamma^a_\mu(z)=\sum_{l=0}^3e^{-i\pi\mu l/2}\gamma^a_l(z),\qquad a\in\{x,y,z\}.
\end{equation}
Then
\be{Wapl}
\mathbf{W}^{(\lambda-\mu)}_{a;+}(w_{n+1})=
-\frac{\theta_2(0)\hat\gamma^a_{\lambda-\mu}(w_{n+1})}{8\theta_1(y+w_{n+1})}.
\ee

\section{Scalar products\label{A-SP}}

Scalar products \eqref{SPdef} satisfy a selection rule
\be{Ortho}
\mathbf S^{\nu,\lambda}_{n,m}(\bv|\bw)\cong \delta_{\nu+n,\lambda+m\;(\hspace{-3mm}\mod 2)}\;.
\ee
In the case of free fermions, it follows from \eqref{Ortho} that either $\lambda=\nu$ or $\lambda=\nu+2\hspace{-1mm}\mod 2$.

It is convenient to introduce
\begin{equation}
    \mathbf S^{\nu;\epsilon}_{n,m}(\bv|\bw)=
    \mathbf S^{\nu,\lambda}_{n,m}(\bv|\bw)+(-1)^\epsilon
    \mathbf S^{\nu,\lambda+2}_{n,m}(\bv|\bw), \qquad \epsilon=0,1.
    \label{SPn-1epsdef}
\end{equation}
Due to the selection rule, non-vanishing scalar products $\mathbf S^{\nu;\epsilon}_{n,m}(\bv|\bw)$ occur for either $\lambda=\nu$ for $m=n$ or $\lambda=\nu+1$ for $m=n\pm1$.

We first give an explicit expression for $S^{\nu;\eps}_{n,n}(\bv|\bw)$ (see \cite{SlaZZ21}). Let
\be{S0}
S^\nu(\bv|\bw)=\frac{\left(-e^{-i\pi\nu/2}\theta'_1(0)\right)^n}{\prod_{\substack{a,b=1\\a\ne b}}^nf(v_a,v_b)}
\left(\prod_{k=1}^n \frac{\chi_\nu(w_k)}{a(v_k)\mathcal{V}_k} \right)\frac{\prod_{a>b}^n\theta_2(v_a-v_b)\theta_2(w_a-w_b)}
{\prod_{a=1}^n\prod_{b=1}^n\theta_1(w_a-v_b)}.
\ee
where $\mathcal{V}_k$ is given by \eqref{logder}.
Then
\be{Svu}
S^{\nu;\eps}_{n,n}(\bv|\bw)= \frac{\theta_1(r+x_\eps)}{\theta_1(x_\eps)} S^\nu(\bv|\bw),
\ee
where
\be{rvu}
r=\sum_{j=1}^n(v_j-w_j).
\ee

Scalar products $S^{\nu;\eps}_{n,n\pm1}$ can be expressed in terms of $S^{\nu;\eps}_{n,n}$. Let $\bw=\{w_1,\dots,w_{n-1}\}$. Then
\begin{equation}
    \mathbf S^{\nu;\epsilon}_{n,n-1}(\bv|\bw)
    = -(-i)^\epsilon
    \frac{\theta_2(0)\theta_1(x_\epsilon)}{2T_\nu(-y^\ast|\bv)}
    \mathbf S^{\nu;\epsilon}_{n,n}(\bv|\{\bw,-y^\ast\}),
    \label{SPn-1eps}
\end{equation}
where $y^*=y-1/2$.

Let now $\bw=\{w_1,\dots,w_{n+1}\}$. Then
\begin{equation}
    \mathbf S^{\nu;\epsilon}_{n,n+1}(\bv|\bw)=\sum_{a>b}^{n+2}
   \frac{ -2(-i)^\epsilon\theta_1(x_\epsilon)\omega_{ab}(-y^\ast)}{\theta_1^2(x)\theta_2^2(x)\theta_2(0)T_\nu(-y^\ast|\bv)}
    \theta_1(w'_a-t_\epsilon)
    \theta_1(w'_b-t_\epsilon)
    \mathbf{S}^{\nu;\epsilon}_{n,n}(\bv|\bw'_{a,b}),
    \label{SPn+1eps_dprod}
\end{equation}
where $\bw'=\{\bw,-y^*\}$ and the coefficients $\omega_{ab}$ are given by \eqref{omg-dbl}.



\begin{thebibliography}{99}
%
\bibitem{KulS23b} G.~Kulkarni, N.~A.~Slavnov, \textsl{Scalar products of Bethe vectors in the generalized algebraic Bethe ansatz},
Theor. Math. Phys. to appear, \texttt{arXiv:2306.12932}.
%
\bibitem{FT79} L. Takhtajan and L. Faddeev, {\sl The quantum method of
the inverse problem and the Heisenberg
$XYZ$ model}, 
Russ. Math. Surveys {\bf 34} (1979), no. 5 11--68.
%
\bibitem{Hei28} W. Heisenberg,
\textsl{Zur Theorie des Ferromagnetismus},
Zeitschrift f\"ur Physik, {\bf 49} (1928) 619--636.
%
\bibitem{Sut70} B. Sutherland, \textsl{Two-dimensional
hydrogen bonded crystals without the ice rule}, J. Math. Phys. {\bf 11} (1970) 3183--3186.
%
\bibitem{FanW70} C. Fan and F. Y. Wu,
\textsl{General lattice model of phase transitions}, Phys. Rev. B {\bf 2} (1970) 723--733.
%
\bibitem{Baxter71}
R. Baxter, {\sl Eight-vertex model in lattice statistics},
Phys. Rev. Lett. {\bf 26} (1971) 832--833.
%
\bibitem{Baxter-book} R. Baxter, {\sl Exactly solved models in statistical mechanics},
Academic Press, 1982.
%
\bibitem{FadST79} L.D. Faddeev, E.K. Sklyanin and L.A. Takhtajan,
\textsl{Quantum Inverse Problem. I},
Theor. Math. Phys. {\bf 40} (1979) 688--706.
%
\bibitem{FadLH96} L.D. Faddeev, \textsl{How Algebraic Bethe Ansatz works
for integrable model}, in: Les Houches Lectures \textsl{Quantum Symmetries}, eds A. Connes
et al, North Holland, (1998) 149, \texttt{arXiv:hep-th/9605187}.
%
\bibitem{BogIK93L}V.E. Korepin, N.M. Bogoliubov,
A.G. Izergin, \textsl{Quantum Inverse Scattering Method and
Correlation Functions}, Cambridge: Cambridge Univ.
Press, 1993.
%
\bibitem{Sla22L} N.A. Slavnov, \textsl{Algebraic Bethe Ansatz and Correlation Functions}, World Scientific, Singapore, 2022.
%
\bibitem{KitMT99} N. Kitanine, J.M. Maillet and V. Terras,
{\sl Form factors of the XXZ Heisenberg spin-$1/2$ finite chain},
Nucl. Phys.  {\bf B554}  (1999) 647--678, \texttt{arXiv:math-ph/9807020}.
%
\bibitem{GohK00} F.G\"ohmann and V.E. Korepin \textsl{Solution of
the quantum inverse problem}, J. Phys. A {\bf 33} (2000) 1199--1220,
\texttt{arXiv:hep-th/9910253}.
%
\bibitem{MaiT00} J.M. Maillet and V. Terras, \textsl{On the quantum inverse scattering problem},
 Nucl. Phys. {\bf B575} (2000) 627--644, \texttt{arXiv:hep-th/9911030}.
%
\bibitem{KulS23} G.~Kulkarni, N.~A.~Slavnov, \textsl{Action of the monodromy matrix entries in the generalized algebraic  Bethe ansatz},
Theor. Math. Phys. to appear, \texttt{arXiv:2303.02439}.
%
\bibitem{SlaZZ21} N. Slavnov, A. Zabrodin, A. Zotov, \textsl{Scalar products of Bethe vectors in the 8-vertex model}, JHEP, {\bf2020}:6 (2020), 123,
\texttt{arXiv:2005.11224}.
%
\bibitem{Sla89}
N.A. Slavnov, {\sl Calculation of scalar products of wave functions and form factors in the
framework of the algebraic Bethe Ansatz}, Theor. Math. Phys. {\bf 79} (1989) 502--508.
%
\bibitem{BS19}
S. Belliard and N. Slavnov, \textsl{Why scalar products in the algebraic Bethe an\-satz
have determinant representation},  J. High Energy Phys. {\bf 10} (2019) 103,
\texttt{arXiv:1908.00032}.
%
\bibitem{LieSM61} E.~Lieb~E, T.~Schultz, D.~Mattis, \textsl{Two soluble models of an antiferromagnetic chain}, Ann. Phys.
{\bf16} (1961) 407--466.
%
\bibitem{Nie67} Th.~Niemeijer, \textsl{Some exact calculations on a chain of spins 12}, Physica {\bf36}:3 (1967) 377--419.
%
\bibitem{McC68}B.~M.~McCoy, \textsl{Spin Correlation Functions of the $X-Y$ Model}, Phys. Rev. {\bf173} (1968) 531--541.
%
\bibitem{KatHS70} S.~Katsura, T.~Horiguchi, M.~Suzuki, \textsl{Dynamical properties of the isotropic $XY$ model},
Physica {\bf46}:1 (1970) 67--86.
%
\bibitem{PerC77}  J.~H.~H.~Perk,  H.~W.~Capel, \textsl{Time-dependent $xx$-correlation functions in the one-dimensional $XY$-model},
Physica A {\bf89} (1977) 265--303.
%
\bibitem{VaiT78a}H.~G.~Vaidya, C.~A.~Tracy, \textsl{Crossover scaling function for the one-dimensional $XY$ model at zero
temperature}, Phys. Lett. A {\bf68} (1978) 378--380.
%
\bibitem{Ton81} T.~Tonegawa, \textsl{Transverse spin correlation function of the one-dimensional spin-$12$ $XY$ model},
Solid State Comm. {\bf40}:11 (1981) 983--986.
%
\bibitem{DlorGS83} M.~D'lorio, U.~Glaus, E.~Stoll, \textsl{Transverse spin dynamics of a one-dimensional $XY$ system:
A fit to spin-spin relaxation data}, Solid State Comm. {\bf47}:5 (1983) 313--315.
%
\bibitem{IzeKS98} A.~G.~Izergin, N.~A.~Kitanin,  N.~A.~Slavnov, \textsl{On correlation functions
of the $XY$ model}, J. Math. Sci. {\bf88}:2 (1998) 224--232.
%
\bibitem{FMC03} K. Fabricius and B. McCoy,
\textsl{New Developments in the Eight Vertex Model}, J. Stat. Phys.,
{\bf 111} (2003) 323--337, \texttt{arXiv:cond-mat/0207177}.
%
\bibitem{FabM05} K. Fabricius and B.M. McCoy,
\textsl{New developments in the eight vertex model II. Chains of odd length}, J. Stat. Phys.
{\bf 120} (2005) 37--70, \texttt{arXiv:cond-mat/0410113}.
\bibitem{FabM04} K. Fabricius and B. M. McCoy, \textsl{Functional Equations and Fusion Matrices for the Eight-Vertex Model}, Publ. RIMS, {\bf 40}  (2004)
905--932, \texttt{arXiv:cond-mat/0311122}.
%
\bibitem{FabM06} K. Fabricius and Barry McCoy, \textsl{An elliptic current operator for the 8 vertex model} J. Phys. A: Math. Gen. {\bf 39} (2006)
14869--14886, \texttt{arXiv:cond-mat/0606190}.
%
\bibitem{Deg02} T. Deguchi, \textsl{The 8V CSOS model and the $sl_2$ loop algebra symmetry of the
six vertex model at roots of unity}, Int. J. Mod. Phys. {\bf B16} (2002) 1899--1905, \texttt{arXiv:cond-mat/0110121}.
%
\bibitem{Deg02a} T. Deguchi, \textsl{Construction of some missing eigenvectors of the $XYZ$ spin
chain at the discrete coupling constants and the exponentially large spectral
degeneracy of the transfer matrix}, J. Phys. A: Math. Gen.  {\bf 35} (2002) 879--895, \texttt{arXiv:cond-mat/0109078}.
%
\bibitem{Fab07} K. Fabricius, \textsl{A new $Q$-Matrix in the Eight Vertex Model},  J. Phys. A: Math. Theor. {\bf 40} (2007) 4075--4086, \texttt{arXiv:cond-mat/0610481}.
%
%
\bibitem{KZ15} S. Kharchev and A. Zabrodin, \textsl{Theta vocabulary I}, Journal of
Geometry and Physics {\bf 94} (2015) 19--31, \texttt{arXiv:1502.04603}.


\end{thebibliography}
\end{document}